\documentclass[final,1p,times]{elsarticle}

\usepackage{graphicx}%
\usepackage{amsmath,amssymb}%
\usepackage{comment}

\journal{}

\begin{document}

\begin{frontmatter}



\title{Stochastic Description of Dynamical Traps in Human Control}


\author[1]{Vasily Lubashevskiy} \ead{vlubashe@tiu.ac.jp}
\author[2,3]{Ihor Lubashevsky}  \ead{ilubashevskii@hse.ru} 
\author[3]{Namik Gusein-zade} \ead{ngus@mail.ru}

\affiliation[1]{%
			organization={Tokyo International University, Institute for International Strategy},
			addressline={4 Chome-42-31 Higashiikebukuro},
			city={Toshima},
			postcode={170-0013}, 
			state={Tokyo}, 
			country={Japan}}

\affiliation[2]{%
			organization={HSE Tikhonov Moscow Institute of Electronics and Mathematics, Department of Applied Mathematics}, 
			addressline={34 Tallinskaya Ulitsa},
			city={Moscow}, 
			postcode={123458}, 
			country={Russia}}

\affiliation[3]{%
	organization={Pirogov Russian National Research Medical University, Faculty of Biomedicine}, 
	addressline={Ulitsa Ostrovityanova, 1},
	city={Moscow}, 
	postcode={117997}, 
	country={Russia}}

\begin{abstract}
A novel model for dynamical traps in intermittent human control is proposed. It describes probabilistic, step-wise transitions between two modes of a subject's behavior---active and passive phases in controlling an object's dynamics---using an original stochastic differential equation. This equation governs time variations of a special variable, denoted as $\zeta$, between two limit values, $\zeta=0$ and $\zeta=1$. The introduced trap function, $\Omega(\Delta)$, quantifies the subject's perception of the object's deviation from a desired state, thereby determining the relative priority of the two action modes. Notably, these transitions---referred to as the subject's action points---occur before the trap function reaches its limit values, $\Omega(\Delta)=0$ or $\Omega(\Delta)=1$. This characteristic enables the application of the proposed model to describe intermittent human control over real objects.

\end{abstract}


\end{frontmatter}


\section{Introduction}\label{sec1}

The concept of a ``dynamical trap,'' initially introduced in \cite{Lubashevsky1998406},\footnotemark{} extends the idea of a stationary point---a central concept in dynamical systems theory---to a region with blurred boundaries. Each point within this region can be considered a stationary point of the corresponding dynamical system. Thus, the dynamical trap represents a multitude of neutral equilibrium states.
When applied to the context of human perception, particularly from the first-person perspective, the dynamical trap becomes a fundamental component of the formalism due to the bounded capacity of human cognition \cite{ihor2017bookmind}. This limitation is especially evident in the inability of humans to clearly distinguish between similar states of an observed object, which differ in terms of certain quantitative parameters in the real world.
The dynamical trap formalism has proven useful in explaining human intermittent control, as demonstrated in experiments involving virtual pendulum balance \cite{zgonnikov2014react} and car-following scenarios in virtual driving \cite{Lubashevsky2018}. Furthermore, dynamical traps can lead to a new class of nonequilibrium phase transitions, where the emergence of new phases is not due to the appearance of new stationary points in the governing equations \cite[for a review]{lubashevsky2016human}.

\footnotetext{A different type of dynamical traps is found in the theory of Hamiltonian systems with complex dynamics \cite{zaslavsky1995from,Zaslavsky2002292}. In this context, a dynamical trap is a region in the phase space with an exceptionally long residence time. This type of traps, however, is fundamentally distinct from the one under consideration, as its emergence results from a delicate balance of several nonlinear properties in a Hamiltonian system. The dynamical trap discussed in this paper, on the other hand, pertains to the general characteristics of human perception.}

The concept of dynamical traps underpins, explicitly or implicitly, the theory of intermittent human control, for example, in the stabilization of an unstable object, where control repeatedly switches between active and passive phases rather than remaining active throughout the process \cite[][for a review]{Insperger2021}. Nowadays, intermittent human control is considered entirely natural, highly effective, and robust \cite{loram2011human}. In the active phase, the subject’s actions aim at making adjustments or correcting the object motion, whereas in the passive phase, the subject temporarily disengages from active control, allowing the system to evolve autonomously. Transitions between the two phases are usually treated as event-driven phenomena, occurring when control is activated or halted due to the discrepancy between the goal and the actual system state exceeding or falling below a certain threshold. These transitions, especially control activation events, can exhibit a pronounced probabilistic behavior, allowing them to be treated as noise-driven control activation \cite{zgonnikov2014react}.
It is essential that in intermittent human control, the active phase fragments should be characterized as ballistic, i.e., open-loop fragments, rather than feedback control \cite{loram2002human} (see also \cite{loram2011human} for a review). In other words, the subject, responding to a critical deviation of the controlled object from the desired state, initiates corrective actions, which are mainly implemented without feedback on the induced variations in the system state. In this sense, the commonality between passive and active phases is that during both, the subject does not \textit{react} to the system dynamics; they simply wait until either the system deviation becomes critical or the action is completed. In this case, exactly the transitions between the two control phases---action points \cite{todosiev1963action} (see also \cite{pariota2015})---should be characterized as real ``active'' actions of the subject caused by event-driven decision-making.

All the proposed models for the dynamical trap effect, which can be categorized under its continuous description, employ the following trap function
\begin{equation}\label{eq:Omega}
	\Omega(\Delta) = \frac{\Delta^2}{\Delta^2 + \Delta_c^2}
\end{equation}  
or a similar function. In general, the variable $\Delta$ represents the intensity of an external stimulus that prompts the subject to act, For instance, $\Delta$ may represent the  deviation of a controlled object from its desired equilibrium position  or the discrepancy between the object's actual motion and its expected dynamics.  The parameter $\Delta_c$ accounts for the uncertainly in the subject's perception of  $\Delta$. Within this approach, the trap function $\Omega(\Delta)$ characterizes the trap effect as a formal reduction in the subject's response to the external stimulus $\Delta$ when $\Delta\lesssim \Delta_c$, approaching zero as $\Delta/\Delta_c \to 0$.

The purpose of the present paper is to propose a stochastic description of the dynamical trap effect that accounts for two characteristic features of human control, which cannot be captured within a continuous approach. First, transitions between active subject's control and temporary disengagement from it typically occur in a step-wise manner, i.e., on time scales determined by the neural-muscular implementation of initiated actions. Second, these transitions must be probabilistic. In this context, the function $\Omega(\Delta)$ should serve as a control parameter that triggers the transitions rather than explicitly describing the subject's actions. In other words, the proposed description should focus on action points in the subject's behavior rather than on smooth transitions between the two modes of action. It should be noted that the model introduced in Ref.~\cite{Zgonnikov2015} may be regarded as a first step toward this approach.

\section{Model}

Two distinct modes of the subject's actions are represented by the limit values $\zeta=0$ and $\zeta = 1$ of a continuous variable $\zeta\in [0,1]$. Specifically, we assume that $\zeta=0$ corresponds to  the subject's disengagement from active control, while $\zeta=1$ characterizes continuous feedback to changes in system dynamics. Relatively sharp transitions of $\zeta$ between 0 and 1 represent action points. Additionally, small fluctuations, $\delta z|_{\zeta=0}>0$ and $\delta z|_{\zeta=1}<0$, near these limit values serve as a quantitative measure of the subject's readiness to switch modes when their preference is no longer clear.

A function $\Omega(t)\in (0,1)$, defined, for example, by Exp.~\eqref{eq:Omega} in terms of the time dependence of the variable $\Delta(t)$, quantifies the proximity of system dynamics to situations where (\textit{i}) corrective intervention by the subject becomes essential or (\textit{ii}) continued active behavior is not only unnecessary but may also cause undesired effects due to the bounded capacity of human cognition. In this study, which focuses on $\zeta$-dynamics, we treat $\Omega(t)$ as a given function of time $t$.    

Since halting and activating active control are considered purely probabilistic phenomena, the limit states $\zeta = 0, 1$ in $\zeta$-dynamics must be metastable. The stability regions of the states $\zeta = 0,1$, which form the interval $(0,1)$, expand or shrink, respectively, as $\Omega$ approaches 0 or 1. In particular, as $\Omega\to 0,1$ the points $\zeta=1,0$ lose their stability, respectively.    

\begin{figure}
\begin{center}
	\includegraphics[width=\textwidth]{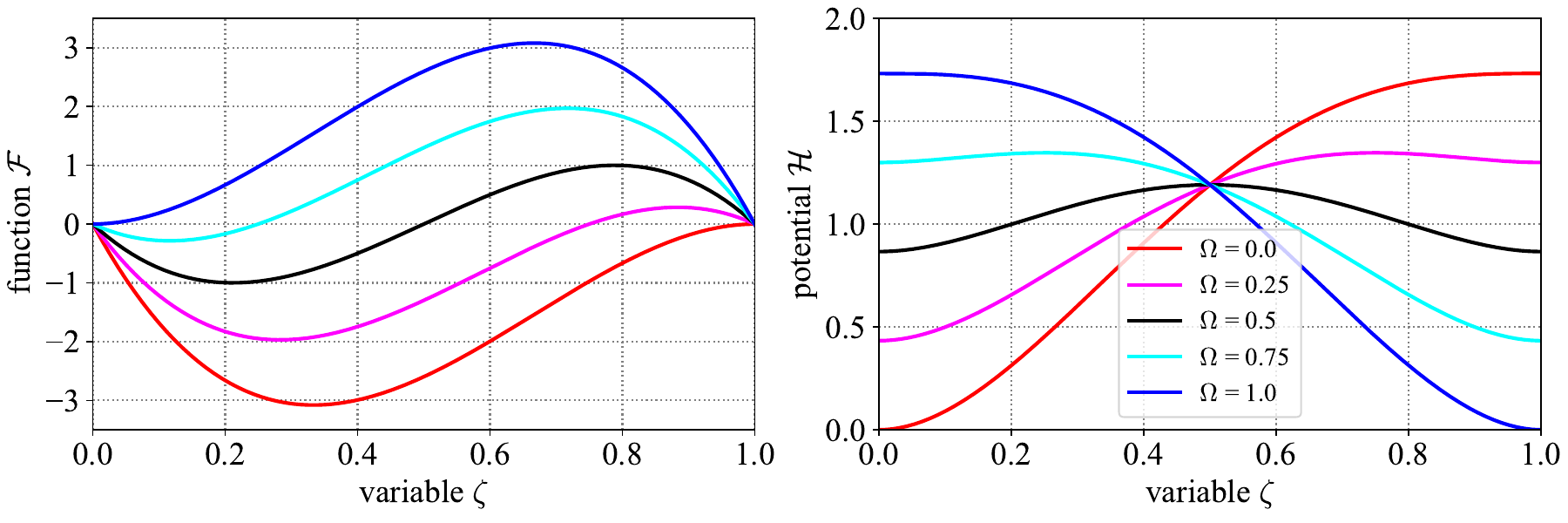}
\end{center}
\caption{Dependence of the function $\mathcal{F}(\zeta,\Omega)$ and the potential $\mathcal{H}(\zeta,\Omega)$ on the variable $\zeta$ for several values of $\Omega$.}	
\label{F:1}	
\end{figure}

This type of $\zeta$-dynamics can be described by the following equation: 
\begin{equation}\label{eq:1}
	\tau_\zeta \frac{d\zeta}{dt} = \mathcal{F}(\zeta,\Omega) +  \tilde{f}(t,\zeta,\Omega)\,,
\end{equation}
where the time scale $\tau_\zeta$ characterizes the neurophysiological delay in the subject's response to changes in system states, the function $\mathcal{F}(\zeta,\Omega)$ is specified by the expression
\begin{equation}\label{eq:1f}
	\mathcal{F}(\zeta,\Omega) = - 12\sqrt{3}\zeta \big(\zeta - 1 + \Omega\big)\big(\zeta - 1\big)\,,
\end{equation}
and the stochastic term $\tilde{f}(t,\zeta,\Omega)$, such that 
\begin{equation}\label{eq:2}
 	\tilde{f}(t,\zeta,\Omega)\big|_{\zeta = 0} \geq 0 \quad\text{and}\quad \tilde{f}(t,\zeta,\Omega)\big|_{\zeta = 1} \leq 0\,,
\end{equation}
quantifies the effect of uncertainty in the subject's perception of system states on decision-making when switching between modes of action. Possible correlations of random fluctuations $\tilde{f}(t,\zeta,\Omega)$ may occur on time scales shorter than $\tau_\zeta$, allowing for the change-of-mind effect during action execution \cite[][for a review]{Stone2022}. The numeric coefficient $12\sqrt{3}$ in Exp.~\eqref{eq:1f} has been selected so that, for $\Omega=1/2$,  the function $\mathcal{F}(\zeta,\Omega)$ attains its maximal value equal to unity (Fig.~\ref{F:1}). 

To ensure that the random term $\tilde{f}(t,\zeta,\Omega)$ possesses the desired properties, we, first, employ the Ornstein-Uhlenbeck process $w(t)$ governed by the equation:
\begin{equation}\label{eq:3}
	\tau_f \frac{dw}{dt} = -w + \sqrt{2\tau_f}   \xi(t)\,,
\end{equation}
where the time scale $\tau_f\lesssim \tau_\zeta$ characterizes the correlations in $\tilde{f}$ and $\xi(t)$ is unit-amplitude white noise, and set
\begin{equation}\label{eq:4}
	\tilde{f}(t,\zeta,\Omega) \propto \big[w(t)\big]^2\,.
\end{equation}
The solution of Eq.~\eqref{eq:3} is given by
\begin{equation}\label{eq:5}
	w(t)=\sqrt{\frac{2}{\tau}}\int\limits_{-\infty}^{t}dt'e^{-(t-t')/\tau}\xi(t')\,,
\end{equation}
which directly yields the following statistical property of the process $[w(t)]^2$:
\begin{equation}\label{eq:6}
	\Big\langle w^{2}(t)w^{2}(t+\delta t)\Big\rangle =1+2e^{-2|\delta t|/\tau_f}.
\end{equation}

Second, the amplitude of random fluctuations $\tilde{f}(t,\zeta,\Omega)$ should increase as the corresponding stability region $(0, 1-\Omega)$ or $(1-\Omega, 1)$ shrinks. To account for this, we write 
\begin{equation}\label{eq:7}
    \tilde{f}(t,\zeta,\Omega) \propto \big[\mathcal{H}(\zeta,\Omega)\big]^{p}\,,
\end{equation}
where the exponent $p>0$ and the potential $\mathcal{H}(\zeta,\Omega)$ shapes the function $\mathcal{F}(\zeta,\Omega)$ as
\begin{equation}\label{eq:8b}
	\mathcal{F}(\zeta,\Omega) = -\frac{\partial \mathcal{H}(\zeta,\Omega)}{\partial\zeta}\,,
\end{equation}
and, additionally, obeys the conditions
\begin{align}\label{eq:9}
	\mathcal{H}(\zeta,\Omega) & > 0\text{ for $0<\Omega<1$},  & \mathcal{H}(0,\Omega) &= 0\text{ for $\Omega=0$}, & \mathcal{H}(1,\Omega) &= 0 \text{ for $\Omega=1$}. 
\end{align}
In other words, the potential $\mathcal{H}(\zeta,\Omega)$ quantifies the relative priority of the states $\zeta=0,1$. Moreover, in the limit cases $\Omega=0$ and $\Omega=1$, the states $\zeta=0$ and $\zeta=1$, respectively, become completely dominant. The potential $\mathcal{H}(\zeta,\Omega)$ meeting these conditions is given by
\begin{equation}\label{eq:8a}
	\mathcal{H}(\zeta,\Omega)  = \sqrt{3}\Big[3\zeta^4 - 4(2-\Omega)\zeta^3 + 6(1-\Omega)\zeta^2 + \Omega \Big],
\end{equation}
as illustrated in Fig.~\ref{F:1}.
Below, we set the exponent $p = 1/2$, since on time scales much longer than $\tau_f$, extreme fluctuations in $ \tilde{f}(t,\zeta,\Omega)$ can be treated as extreme fluctuations of white noise, by virtue of \eqref{eq:8a}, and exactly the quantity $\big[\tilde{f}(t,\zeta,\Omega)\big]^2$, together with $\mathcal{F}(\zeta,\Omega)$, determines escape from the corresponding potential well. 

Finally, we introduce the cofactor $\cos(\pi \zeta)$ to ensure that inequalities~\eqref{eq:2} are satisfied. As a result, the proposed model for the uncertainty in the subject's actions is given by the following random term:
\begin{align}\label{eq:10}
	\tilde{f}(t,\zeta,\Omega) = \epsilon \sqrt{\mathcal{H}(\zeta,\Omega)} \cos(\pi\zeta) \big[w(t)\big]^2\,, 
\end{align}
where $\epsilon\ll 1$ is a model parameter.

\section{Numerical simulation}

To analyze the probabilistic properties of action points, represented by sharp transitions between the states $\zeta = 0$ and $\zeta = 1$ in $\zeta$-dynamics, the developed model rewritten in the dimensionless form
\begin{subequations}\label{sim:1}
\begin{align}
\label{sim:1a}
	\frac{d\zeta}{dt} & = \mathcal{F}[\zeta,\Omega(t)] +  \epsilon \sqrt{\mathcal{H}[\zeta,\Omega(t)]} \cos(\pi\zeta) w^2,\\
\label{sim:1b}
	\frac{dw}{dt}     & = -\rho w + \sqrt{2\rho}\xi(t)\,, \\
\label{sim:1c}
	\Omega(t)         & = \frac12 \big[ 1 + \Lambda \sin\left({t}/{S_t}\right)\big]\,,
\end{align}
\end{subequations}
has been studied numerically. Here the parameter $\rho = \tau_\zeta/\tau_f\gtrsim 1$, the time dependence \eqref{sim:1c} mimics the dynamics of a controlled object, in which the subject has to switch regularly between the two modes of action on a characteristic (dimensionless) time scale $S_t$. Initially, we may assume $S_t \gtrsim 1$ as on time scales comparable to $\tau_\zeta$ the subject is unable to effectively control the object motion. The parameter $0 < \Lambda < 1$ accounts for the fact that the subject must respond to changes in object motion when $\Delta \sim \Delta_c$ (Eq.~\ref{eq:Omega}), rather than in extreme cases where $\Delta \gg \Delta_c$ or $\Delta \ll \Delta_c$.

System~\eqref{sim:1} was integrated using the strong stochastic Runge-Kutta method SRI2W1 of order 1.5 for stochastic differential equations with scalar noise \cite{Rossler_2010}. The total integration time was $T=10^6$, the time step was set to $0.001$, and the parameters $\epsilon$, $\rho$, and $\Lambda$  were fixed at $\epsilon=0.15$, $\rho=3$, and $\Lambda = 0.6$.

\begin{figure}
	\begin{center}
		\includegraphics[width=\textwidth]{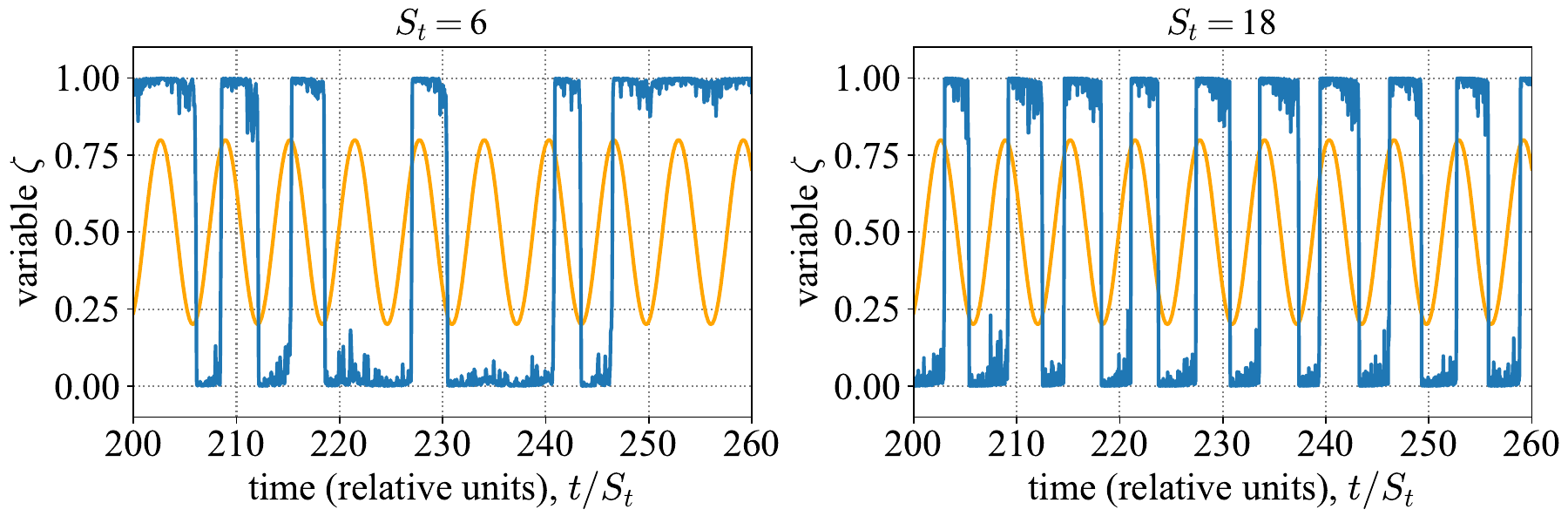}
	\end{center}
	\caption{Characteristic time patterns of $\zeta$-dynamics for different values of $S_t$. The orange lines depict the time dependence of $\Omega(t)$, determined by Exp.~\eqref{sim:1c} with $\Lambda = 0.6$.}	
	\label{F:2}	
\end{figure}

\begin{figure}
	\begin{center}
		\includegraphics[width=\textwidth]{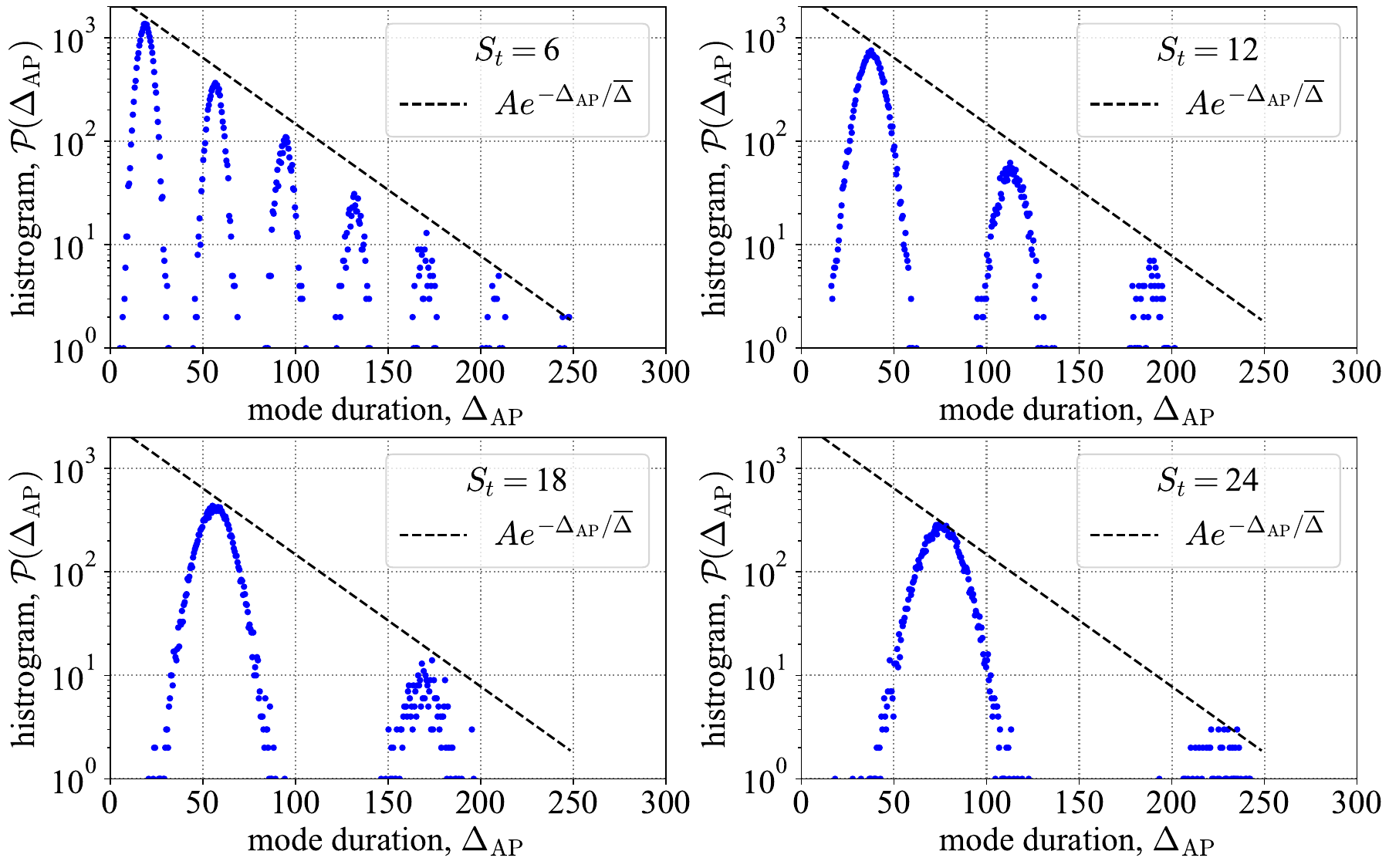}
	\end{center}
	\caption{Histograms of time intervals between successive action points, representing the duration of single-mode fragments in $\zeta$-dynamics, for $\Omega(t)$ as defined by Exp.~\eqref{sim:1c}. The specific values of $A$ and $\overline{\Delta}$ depend on the model parameters $\epsilon$, $\rho$, and $\Lambda$ fixed in the presented simulation. Notably, $A$ is also proportional to the total integration time $T$. In this case, $A=2800$ and $\overline{\Delta} = 34$.}	
	\label{F:3}	
\end{figure}

Figures~\ref{F:2} and \ref{F:3} present the simulation results. In particular, Figure~\ref{F:2} depicts the characteristic patterns of $\zeta$-dynamics for two values of $S_t$, namely 6 and 18. These patterns can be interpreted as a sequence of stepwise transitions between the states $\zeta=0$ and $\zeta=1$, corresponding to instances where the subject takes action to change the system's control mode.
These patterns illustrate that the dynamical trap function $\Omega$, given fixed system parameters $\epsilon$ and $\rho$, determines the characteristic time $\overline{T}_\Omega$ for the occurrence of action points. In other words, $\overline{T}_\Omega$ represents the typical duration within which the subject decides to switch the current mode of behavior to optimize control over the object motion. Mathematically, this decision-making process corresponds to an escape from a potential well.
For $S_t=6$, the values of $\overline{T}_{\Omega=0.2}$ and $\overline{T}_{\Omega=0.8}$ significantly exceed $S_t$, preventing transitions between the states $\zeta=0$ and $\zeta=1$ each time $\Omega(t)$ reaches its extreme values, $\Omega_{\min} = 0.2$ and $\Omega_{\max} = 0.8$. As shown in Figure~\ref{F:2}, when $S_t=18$, the time scales $\overline{T}_{\Omega=0.2}$, $\overline{T}_{\Omega=0.8}$, and $S_t$ become comparable, facilitating more frequent state transitions.

Figure~\ref{F:3} supports the proposition regarding the time scale $\overline{T}_{\Omega}$. It presents histograms of the time intervals $\Delta_\text{AP}$ between successive action points (0$\leftrightarrow$1-transitions) for various values of $S_t$. Due to the chosen form of $\Omega(t)$ (Exp.~\ref{sim:1c}), these histograms exhibit a series of peaks with diminishing heights. The envelope of these peaks, consistent across all $S_t$ values, defines the probability distribution $\mathcal{P}(\Delta_\text{AP})$ for the time intervals between action points, representing the duration of single-mode fragments. As evident from Figure~\ref{F:3}, this distribution follows a Laplace form:
\begin{equation}\label{eq:Lap} 
	\mathcal{P}(\Delta_\text{AP}) \propto \exp\left(- \frac{\Delta_\text{AP}}{\overline{\Delta}} \right), 
\end{equation}
which provides an estimate of $\overline{T}_{\Omega=0.2} = \overline{\Delta} \approx 34$.

\section{Conclusion}

We have developed a dynamical trap model that describes probabilistic, step-wise transitions between two modes of subject's behavior. These transitions can be interpreted as action points, where the subject decides to change the current mode. One mode represents continuous control over an object's dynamics, while the other involves temporary disengagement from the subject's feedback to variations in the object's motion. Notably, the latter mode encompasses both cessation of the subject's actions and ballistic implementations of initiated corrections to the object's dynamics.

The introduced trap function $\Omega(\Delta)$ quantifies uncertainty in the subject's perception of the object's deviation ($\Delta$) from a desired state. This function serves as a measure of the necessity for the subject's feedback to temporal variations in the object's motion. In other words, $\Omega(\Delta)$ quantifies the priority of the two modes of the subject's behavior. It is worth noting that these transitions can occur much earlier than $\Omega(\Delta)$ approaches its limit values, i.e., when $\Omega(\Delta) \ll 1$ or $1 - \Omega(\Delta) \ll 1$. This feature opens the door to applying the proposed model to describe real processes of intermittent human control.

The found distribution of single-mode fragments---time intervals between successive action points---follows the Laplace distribution, which is commonly observed in human actions when controlling the dynamics of various physical objects \cite{ihor2017bookmind,zgonnikov2014react,Lubashevsky2018,lubashevsky2016human}.





\end{document}